\documentclass[multphys,vecphys]{svmult}

\usepackage{makeidx}     
\usepackage{graphicx}    

\makeindex             

\begin{document}

\title*{On the Influence of Uncertainties in Chemical Reaction Rates on
Results of the Astrochemical Modelling}
\titlerunning{Rate Uncertainties and Astrochemical Modelling}
\author{A.I. Vasyunin\inst{1}, A.M. Sobolev\inst{1},
D.S. Wiebe\inst{2} \and D.A. Semenov\inst{3}}


\institute{Ural State University, Ekaterinburg, Russia
\texttt{Andrej.Sobolev@usu.ru} \and Institute of Astronomy of the
RAS, Moscow, Russia \texttt{dwiebe@inasan.rssi.ru} \and
Astrophysical Institute, Jena University, Germany
\texttt{dima@astro.uni-jena.de}}
%
%
\maketitle


{\textbf {\center Abstract.}} With the chemical reaction rate
database UMIST95 (Millar etal. 1997) we analyze how uncertainties
in rate constants of gas-phase chemical reactions influence the
modelling of the molecular abundances in the interstellar medium.
Random variations are introduced into the rate constants to
estimate the scatter in theoretical abundances. Calculations were
performed for the dark and translucent molecular clouds where gas
phase chemistry is adequate (Terzieva \& Herbst 1998). Similar
approach was used by Pineau des Forets \& Roueff (2000) for the
study of chemical bistability. All the species are divided into 6
sensitivity groups according to the value of the scatter in their
model abundances computed with varied rate constants. It is shown
that the distribution of species within these groups depends on
the number of atoms in them and on the adopted physical
conditions. The simple method is suggested which allows to single
out reactions that are most important for the evolution of a given
species.

\section{Scatter in the model abundances}

To study the influence of uncertainties in reaction rates on the
model abundances we calculated $10^4$ variants of each species
abundance. Figure 1 shows the scatter for some representative
species.

\begin{figure}
\centering
\includegraphics[height=10cm,clip=]{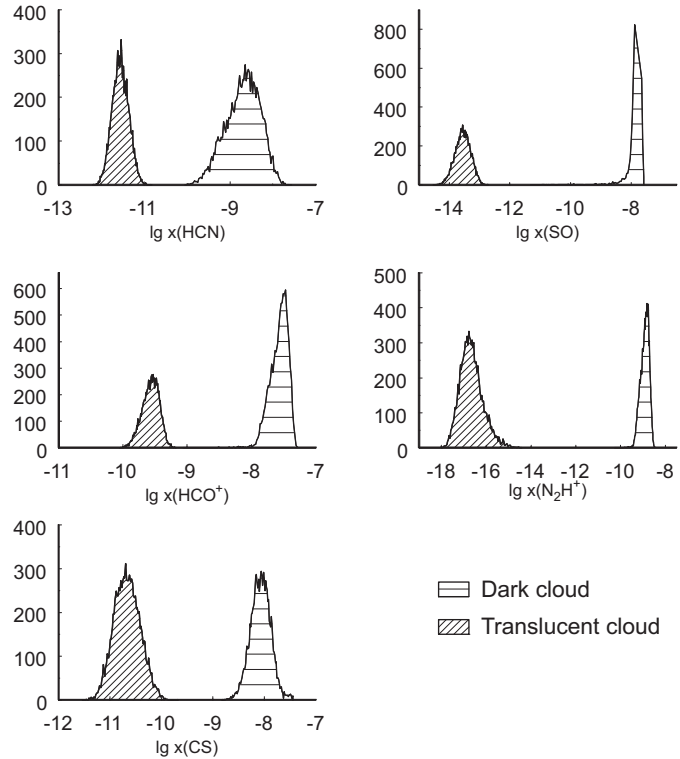}
\caption{The distribution of 10000 model abundances calculated
with different rate coefficients} \label{fig:1}
\end{figure}

We divided all species into 6 sensitivity groups according to the
value of the scatter in their model abundances computed with
varied rate constants.

Distribution of the species within these groups depends on adopted
physical conditions (see fig. 1). Scatter in logarithmic
abundances of simple molecules lies within $0.5 - 1$~dex. Figure 2
shows how the species are distributed by the number of atoms in
different sensitivity groups. It is clear that the scatter in
abundances significantly increases with the number of atoms in the
molecule.

\begin{figure}
\centering
\includegraphics[height=10cm]{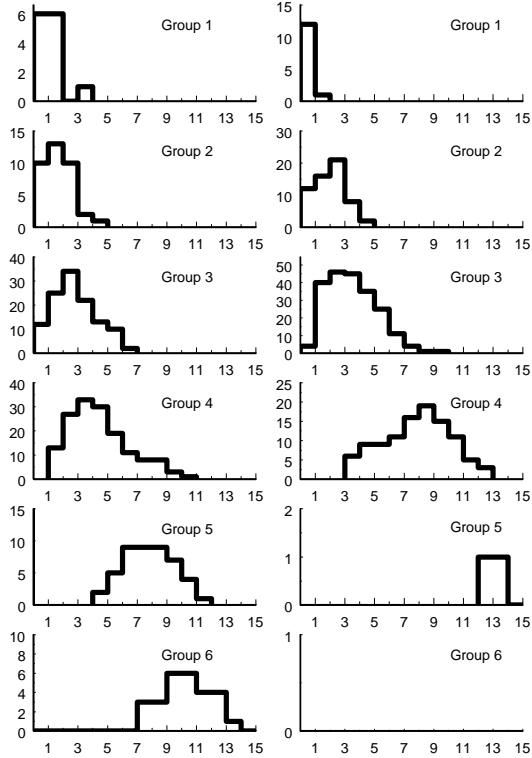}
\caption{Distribution of the molecules by number of atoms for
different sensitivity groups. Left column -- dark cloud, right
column -- translucent cloud} \label{fig:2}
\end{figure}

As an example we considered two molecules, HCN and C$_2$S, which
in the gas-phase chemistry reach maximum of their abundance at
about $10^4$~years and, hence, considered to be characteristic
representatives of the ``early" chemistry. For example, high
abundances of HC$_3$N and other cyanopolyynes in the dark cloud
TMC-1 are considered to prove that this object is chemically young
(Hirahara etal. 1992).

Our calculations show that in the dark cloud model HC$_3$N and
C$_2$S molecules belong to the sensitivity group 4, while more
complex cyano\-polyynes (HC$_5$N, etc.) fall in the groups 5 and
6.

Left panel of fig.3 displays time dependence of the HC$_3$N
abundance calculated with the sets of constants providing maximal,
minimal and 2 intermediate equilibrium abundances. It is clear
that variations of the equilibrium HC$_3$N abundance are
accompanied by noticeable changes of the time when the maximal
abundance is achieved (from 7$\cdot$10$^4$ to 3$\cdot$10$^5$ yr).
The right panel displays corresponding dependence for C$_2$S. The
peak in the C$_2$S abundance time dependence is more flat and
shifts less. Anyhow, quantitative estimates of the molecular cloud
ages on the basis of molecular abundances and their ratios (as
suggested in, e.g., in Stahler 1984) are subject to this
additional uncertainty.

\begin{figure}
\centering
\includegraphics[height=4cm]{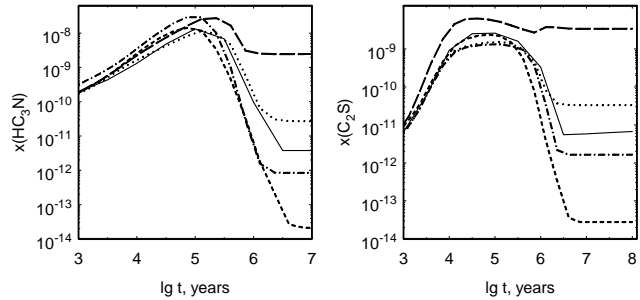}
\caption{Evolution of the HC$_3$N and C$_2$S abundances calculated
with different rate coefficients}
\label{fig:3}       
\end{figure}

\section{Correlation between the species abundance and reaction rate constant}

Because of nonlinearity of the rate equations it is impossible to
sort out analytically the reactions which rate uncertainty greatly
affects the total uncertainty of a species abundance. We propose a
simple statistical method to resolve the issue. The method is
based on investigation of correlations between the species
abundance and rate constants of the reactions and allows to
estimate how the increased precision of the reaction rate constant
will decrease the uncertainty in the species abundance.

Figure 4 shows distribution of HCO$^+$ abundances in the plane
``rate coefficient -- abundance". The left panel represents the
coverage of the plane for reaction with the large correlation
coefficient H$_2$ + C.R.P. $\rightarrow$ H$_2$$^+$ + e$^-$
($r=0.84$) and the right panel -- that for reaction with the small
coefficient C~+~NH~$\rightarrow$~CH~+~N ($r=0.01$). Additional
modelling makes possible to predict improvement of the abundance
estimate for the case when rates of several reactions are going to
be remeasured.


\begin{figure}[h!]
\centering
\includegraphics[height=4.4cm]{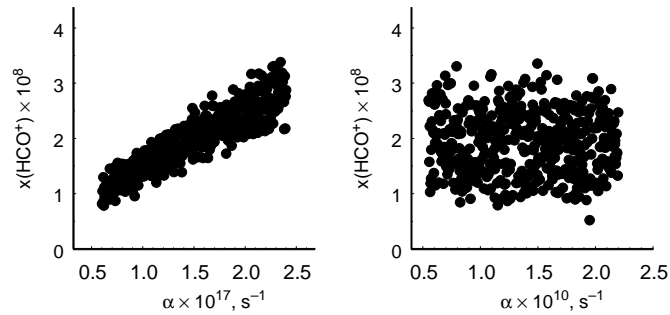}
\caption{HCO$^+$ model abundances in the ``rate
coefficient-abundance" plane}
\label{fig:4}       
\end{figure}

{\bf Acknowledgements.} AIV and AMS were supported by the grants
of INTAS, RFBR and Russian Ministry of Education. DSW was
supported by the RFBR grants 01-02-16206 and 02-02-04008 and by
the President of the RF grant MK-348.2003.02.

%
%
%

\end{document}